\documentstyle[12pt,aaspp]{article}
\begin{document}
\title{INTEGRAL SOLUTION FOR THE MICROWAVE 
BACKGROUND ANISOTROPIES IN NON-FLAT UNIVERSES}

\author{Matias Zaldarriaga}
\affil{Department of Physics, MIT, Cambridge, MA 02139 USA}
\author{Uro\v s Seljak}
\affil{ Harvard Smithsonian Center For Astrophysics, Cambridge, MA 02138 USA}
\author{Edmund Bertschinger}
\affil{Department of Physics, MIT, Cambridge, MA 02139 USA}
\def\bi#1{\hbox{\boldmath{$#1$}}}

\begin{abstract}
We present an efficient method to compute CMB anisotropies in non-flat 
universes.  First we derive the Boltzmann equation for cosmic microwave 
background temperature and polarization fluctuations produced by scalar 
perturbations in a general Robertson-Walker universe. We then apply the 
integral method to solve this equation, writing temperature and 
polarization anisotropies as a time integral over a geometrical term and
a source term. The geometrical terms can be written using
ultra-spherical Bessel functions, which depend on curvature. These cannot 
be precomputed in advance as in  flat space. Instead we solve directly 
their differential equation for selected values of the multipoles. The 
resulting computational time is comparable to the flat space case and
improves over previous methods by 2-3 orders of magnitude. This allows
one to compute highly accurate CMB temperature and polarization
spectra, matter transfer functions and their CMB normalizations for any 
cosmological model, thereby avoiding the need to use various 
inexact fitting 
formulae that exist in the literature.
\end{abstract}

\keywords{cosmology: cosmic microwave background,
cosmology: large-scale structure
of the universe, gravitation, cosmology: dark matter} 
\newpage

\section{Introduction}

In the past few years  
the field of cosmic microwave background (CMB) 
anisotropies transformed from a theoretical exercise to an active
experimental area of research. Since the first 
discovery by COBE in 1992 (Smoot et al. 1992) there have been more than 
a dozen independent detections of fluctuations over a much larger 
angular range (see e.g. Lineweaver et al. 1996 and Rocha and Hancock
1996 for  compilations).
The future looks even more promising, as there are two funded satellite
missions in planning, 
one in Europe (Planck) and one in USA (MAP), 
that have the promise to measure
the anisotropies to an exquisite accuracy over three decades in 
angular scale. 
This will open the 
possibility of determining many 
cosmological parameters with much greater accuracy than any other probe
in cosmology (\cite{Jungman95}; \cite{Bond97}; \cite{zal97}). 

In light of
this experimental promise theoretical predictions have been 
advanced to a higher stage as well. It is not sufficient anymore
to make qualitative statements and approximate predictions to 
analyze the forthcoming data and to 
guide experimental design. 
Rather what is needed are very accurate predictions 
of CMB spectra (typically to better than 1\%), 
which allow one to study the CMB
sensitivity to various cosmological parameters. Because this sensitivity 
depends on the choice of mission specifics such as number of
detectors, their noise characteristics or angular resolution, it 
is important that such accurate predictions are available 
already at the mission design stage when its characteristics
are still  open to modification. 
In Seljak \& Zaldarriaga
1996 (hereafter Paper I) we presented a method  that is both 
fast, accurate and applicable to any flat
cosmological model. The method is based on the source time integration
over the photon past light cone and
has the advantage of reducing the computational time by about two orders of
magnitude compared to the more traditional methods, while still 
being exact in the sense that it can achieve arbitrary 
precision within the limits of linear perturbation theory. 
Such a method therefore allows one to explore a large range of parameter
space with high accuracy.

Open cosmological models and their predictions for CMB 
have received much attention lately (e.g.
Gouda \& Sugiyama 1992, Kamionkowski, 
Spergel \& Sugiyama 1994, Hu, Bunn \& Sugiyama 1995, White \& Bunn 1995,
Gorski et al. 1995). The reason for this is simple:
observational evidence from the nearby universe (e.g. Peebles 1993) suggests 
that an open universe is favored over the critical or  
closed one. While the differences might be resolved with
the addition of a cosmological constant $\Lambda$, 
so that the geometry of the 
universe would remain flat, the value of $\Lambda$
is already significantly constrained
by a number of independent tests based on COBE data, lensing 
statistics and SN type Ia results 
(Bunn \& White 1997; Kochanek 1996; Perlmutter et al. 1996). It therefore
seems 
natural to explore the  possibility 
that the universe is indeed open. 
   From a theoretical point of view,
open inflationary models proved to be constructible
and capable of generating the initial perturbations
(e.g. Lyth \& Stewart 1990, Ratra \& Peebles 1994, Liddle et al. 1995,
Bucher et al. 1995).  
Models with positive curvature, although currently less popular, 
have also been studied in detail (White \& Scott 1996). 

All the
computations of CMB anisotropies in open and closed universes 
carried out so far have 
used the traditional Boltzmann hierarchy approach pioneered by 
Peebles \& Yu (1970) and extended by Wilson \& Silk (1981) and
Bond \& Efstathiou (1984). 
The speed limitations of the Boltzmann approach in  flat space
were extensively discussed in Paper I. 
In the case of non-flat geometry they become even
worse: solving the 
hierarchy in an open universe is much slower 
than in the
flat case and leads to extremely long integration times. This makes  
an accurate search over a large set of parameter space practically 
impossible. Instead one is forced to use approximations, but as shown
in this paper these
give at best a few percent accuracy and can be very misleading in
the study of parameter determination.  
   
In this paper we develop the integral solution for photon transport
in a general Robertson-Walker background, thereby generalizing 
the method developed 
in Paper I to a non-flat geometry. As in the flat case 
the solution is written in 
terms of a time integral over a source term and a geometrical term. 
The latter can be expressed in terms of
ultra-spherical Bessel functions, defined as  
the radial part of the 
eigenfunction of the Laplacian on a curved manifold. 
We present a method for 
computing these functions efficiently in the context of CMB calculations
and incorporate them into the integral solution. This allows us to 
achieve a fast and accurate method,  
improving over previous calculational methods by 2-3 orders of magnitude.
In this paper 
we concentrate on the temperature and polarization anisotropies
produced by scalar modes, the generalization to vector and tensor
perturbations will be presented elsewhere (\cite{coll}). 
The outline of the paper is as follows: in \S2 we present 
the Einstein and fluid
equations in a general Robertson Walker background, in \S3
we discuss the Boltzmann hierarchy for CMB photons and
in \S4 we derive its integral solution. 
In \S5 we discuss in detail the 
method for computing the ultra-spherical Bessel functions. The
comparison between the exact solution and the
approximations often used in the literature is presented in \S6.
In \S7 we 
discuss 
the numerical implementation of the method. 
This is followed by discussion and conclusions in \S8.
In an appendix we present an alternative derivation of the 
integral solution, highlighting its geometrical interpretation.

\section{Einstein and Fluid Equations}

In this section we present the Einstein and fluid differential equations 
for the metric, cold dark matter (CDM)  and baryons that must be  solved
to calculate the CMB anisotropy spectra. These equations are the 
basis of the traditional methods and are also used in the integral 
method, discussed in \S 4. The derivation of the Einstein and  fluid 
equations 
in a non-flat background can be found in the literature (see for example 
Bertschinger 1996), 
so we just present the final results. 

The metric is written as  
\begin{eqnarray}
ds^2&=&-dt^2+a^2(\gamma_{ij}+h_{ij})dx^i dx^j \nonumber \\
 &=& a^2 [-d\tau^2+(\gamma_{ij}+h_{ij})dx^i dx^j],
\end{eqnarray}
where $a$ is the expansion factor, $x_i$ the comoving coordinates and
$\tau=\int dt /a$ the conformal time. We are using units in which
$c=1$.  Space part of unperturbed metric is $\gamma_{ij}$ with
constant curvature $K=H_0^2(\Omega_0-1)$ and $h_{ij}$ is the metric
perturbation in synchronous gauge (Lifshitz 1946). Although all 
observable quantities are identical in different gauges the   
computational efficiency to obtain them within a given accuracy 
is not. This criterion lead us to 
work in synchronous gauge \footnote{In paper I we presented the
equations in the longitudinal gauge.}. In comparison to  
the longitudinal gauge (\cite{bardeen}) it is about 20\% more efficient with 
isentropic initial conditions and even more so with isocurvature
initial conditions, which are difficult to set up in 
the longitudinal gauge. 

When 
working with linear theory in a flat universe it is convenient to use  
Fourier modes as they evolve independently. Their generalization in a 
non-flat
universe are the eigenfunctions of the Laplacian operator
that we shall call $G(\vec k,\vec x)$ (e.g. Abbott \& Schaefer 1986), 
\begin{equation}
\nabla^2 G(\vec k,\vec x)= -k^2 G(\vec k,\vec x).
\label{Laplace}
\end{equation}  

We expand all the perturbations in terms of $G$ and its spatial 
covariant 
derivatives. For example, the metric perturbations for a single mode are 
given by 
\begin{equation}
h_{ij}={h\over 3}\gamma_{ij}G+
(h+6\eta)(k^{-2}G_{\mid ij}+{1\over 3}\gamma_{ij}
G),
\end{equation}
where 
$h$ and $(h+6\eta)$ are the trace and traceless part of the metric 
perturbation. Covariant derivatives with respect to the metric
$\gamma_{ij}$ are denoted with a ``$|$''
The perturbed Einstein's equations result in the following equations for $h$
and $\eta$ (Bertschinger 1996),
\begin{eqnarray}
(k^2-3K)\eta-{1\over 2}{\dot a \over a} \dot{h}&=&-8\pi {\cal G}
a^2\delta \rho 
\nonumber \\
(k^2-3K) \dot{\eta}+{K\over 2} \dot{h}&=&4\pi {\cal G} a^2 
({\bar\rho}+{\bar p})kv,  
\end{eqnarray}
$\cal G$ here stands for the gravitational constant;
$\delta \rho$ and $v$ characterize the density and velocity 
perturbations 
($v=i \hat{k} \cdot \vec v$), 
$\delta \rho = \sum_j {\bar \rho_j} \delta_j$, 
$({\bar\rho}+{\bar p})v = \sum_j ({\bar\rho_j}+{\bar p_j})v_j$,
where
${\bar\rho_j}$ and ${\bar p_j}$ are the mean density and pressure of 
the $j$-th species
and the sum is carried out over all the different species in the universe.   

The 
equation for the cold dark matter density perturbation $\delta_c$ is,
\begin{equation}
\label{cdm2}
        \dot{\delta_c} = -{\dot h \over 2},
\end{equation}
where by definition in this gauge the cold dark matter particles have
 zero peculiar velocities. The Euler equation for the baryons
has additional terms 
caused by Thomson scattering and  pressure, so baryons  
have velocities relative to the dark matter,
\begin{eqnarray}
\label{baryon2}
\dot{\delta}_b &=& -kv_b-{\dot{h}\over 2} \,, \nonumber\\
\dot{v}_b &=& -{\dot{a}\over a}v_b
+ c_s^2 k\delta_b
+ {4\bar\rho_\gamma \over 3\bar\rho_b}
 an_ex_e\sigma_T(v_{\gamma}-v_b) \,.
 \label{cdmb}
 \end{eqnarray}
Here $c_s$ is the baryon sound speed, $v_b$ is the baryon velocity,
$v_{\gamma}$ is given by the
temperature dipole $v_{\gamma}=3\Delta_{T1}$ and
$\bar\rho_\gamma$, $\bar\rho_b$
are the mean photon and baryon densities, respectively.
The Thomson scattering cross section is $\sigma_T$, 
$n_e$ is the electron density and  $x_e$ is the ionization fraction.

\section{Boltzmann equation}

In this section we discuss CMB anisotropies and
present a derivation of the Boltzmann
equation for the photons. Because we use a novel approach to 
treat this problem (see also Hu \& White 1997) we start 
by considering flat geometries in \S3.1 and then 
generalize the results to arbitrary Robertson-Walker backgrounds in 
\S3.2. The derivation of the Boltzmann hierarchy for polarization 
is new to this work.
The notation we use is largely based on Paper I.

The CMB radiation field is described by a $2\, \times \, 2$ 
intensity tensor
$I_{ij}$
(\cite{chandra}). The Stokes parameters $Q$ and $U$ are defined as 
$Q=(I_{11}-I_{22})/4$ and $U=I_{12}/2$, while the temperature anisotropy 
is given by $T=(I_{11}+I_{22})/4$. The fourth Stokes parameter $V$ that
describes circular polarization is not necessary in standard cosmological 
models because it cannot be generated through the process of Thomson 
scattering. While the temperature is a scalar quantity $Q$ and $U$ are
not. They depend on the direction of observation $\hat n$
and on the two axis $(\hat e_1, \hat e_2)$ 
perpendicular to $\hat n$ used to define them. If for a given 
$\hat n$ the axes $(\hat e_1, \hat e_2)$ are rotated by an angle
$\psi$ such that 
${\hat e_1}^{\prime}=\cos \psi \ {\hat e_1}+\sin\psi \ {\hat e_2}$ 
and ${\hat e_2}^{\prime}=-\sin \psi \ {\hat e_1}+\cos\psi \ {\hat e_2}$
the Stokes parameters change as
\begin{eqnarray}
Q^{\prime}&=&\cos 2\psi \  Q + \sin 2\psi \ U \nonumber \\  
U^{\prime}&=&-\sin 2\psi \ Q + \cos 2\psi \ U
\label{QUtrans} 
\end{eqnarray}

To analize the CMB temperature on the sky it is natural to
expand it in spherical harmonics. These are not appropriate 
for polarization, because   
the two combinations $Q\pm iU$ are quantities of spin $\pm 2$
(\cite{goldberg}). They 
should be expanded in spin-weighted harmonics $\, _{\pm2}Y_l^m$ 
(\cite{spinlong}),
\begin{eqnarray}
T(\hat n)&=&\sum_{lm} a_{T,lm} Y_{lm}(\hat n) \nonumber \\
(Q+iU)(\hat{n})&=&\sum_{lm} 
a_{2,lm}\;_2Y_{lm}(\hat{n}) \nonumber \\
(Q-iU)(\hat{n})&=&\sum_{lm}
a_{-2,lm}\;_{-2}Y_{lm}(\hat{n}).
\label{Pexpansion}
\end{eqnarray}
There is an equivalent expansion using tensors on the
sphere (Kamionkowski, Kosowsky \& Stebbins 1997). 
The coefficients $a_{\pm 2,lm}$ 
are observable on the sky and their power spectra
can be
predicted for different cosmological models. Instead of $a_{\pm 2,lm}$
it is convenient
to use their linear combinations
$a_{E,lm}=-(a_{2,lm}+a_{-2,lm})/2$ and 
$a_{B,lm}=-(a_{2,lm}-a_{-2,lm})/2i$, which have opposite parities. 
Four power spectra are needed to
characterize fluctuations in a gaussian theory,
the autocorrelation of 
$T$, $E$ and $B$ and the cross correlation of $E$ and $T$.
Because of parity considerations the cross-correlations
between $B$ and the
other quantities vanish and one is left with
\begin{eqnarray}
C_{Xl}&=&{1\over 2l+1}\sum_m \langle a_{X,lm}^{*}
a_{X,lm}\rangle 
\nonumber \\
C_{Cl}&=&{1\over 2l+1}\sum_m \langle a_{T,lm}^{*}a_{E,lm}\rangle 
\label{Cls},
\end{eqnarray}
where
$X$ stands for $T$, $E$ or $B$ and $\langle\cdots \rangle$
means ensemble average.  
Only vector and tensor modes contribute to $B$ (\cite{spinlong,kamion96}), 
hence we may ignore it in the remainder of this paper.  

\subsection{Flat geometry}

We will start by studying perturbations in a flat
universe to introduce our notation and clarify the treatment of 
polarization, which differs from the usual method 
where Legendre polynomials are used to expand both 
temperature and polarization (e.g. Bond \& Efstathiou 
1984). 
As we are dealing with a linear problem we may
consider only one eigenmode of the Laplacian (a plane wave in the flat 
case) at a time. 
We may choose without loss of generality that
$\vec k \parallel \hat z$. To define the 
Stokes parameters we use the spherical coordinate unit vectors
$(\hat e_{\theta},\hat e_{\phi})$. In this particular coordinate system
only $Q$ is different from zero and we denote it by
$\Delta_P$, so that $\Delta_P=Q=Q\pm i U$ $(U=0)$. 
The temperature anisotropy for the single eigenmode is denoted
by $\Delta_T$.

For a single plane wave rotational symmetry implies that 
both $\Delta_T$ and $\Delta_P$ depend only on the angle between
$\hat n$ and $\hat z$ ($\vec k \parallel \hat z$),
so only harmonics with $m=0$ are needed in the expansion. To calculate 
 the evolution of these two
quantities, we expand them as
\begin{eqnarray}
\Delta_T(\vec k, \hat n)&=&\ G(\vec k,\vec x)
\sum_l (-i)^l \sqrt{4\pi(2l+1)}\Delta_{Tl} \; Y_l^0\nonumber \\
&=&\ G(\vec k,\vec x)\sum_l (-i)^l (2l+1)\Delta_{Tl} P_l(\mu)
\nonumber \\
\Delta_P&=& \ G(\vec k,\vec x)\sum_l (-i)^l \sqrt{4\pi(2l+1)
(l+2)!/(l-2)!} \; _2\Delta_{Pl}
\; _2Y_l^0(\mu)\nonumber \\
&=& \ G(\vec k,\vec x)\sum_l (-i)^l \sqrt{4\pi(2l+1)
(l+2)!/(l-2)!} \; _{-2}\Delta_{Pl}
\; _{-2}Y_l^0(\mu)  \nonumber \\
&=&\ G(\vec k,\vec x)\sum_l (-i)^l (2l+1)
 \; _2\Delta_{Pl}
P_l^2(\mu) 
\label{expleg}
\end{eqnarray}
where $G (\vec k,\vec x) =\exp(i\vec k \cdot
\vec x)$ and $\mu=\hat k \cdot \hat n$. 
%A similar expansion has been 
%presented recently by Hu \& White (1997).
We added  a subindex $\pm 2$ to $_{\pm 2}\Delta_{Pl}$ to  denote that they
are the expansion coefficients in spin $\pm 2$ harmonics \footnote{The
relation between these coefficient and those used in Zaldarriaga \&
Seljak (1997)
is $_{\pm 2}\Delta_{Pl}= - \sqrt{(l-2)!/(l+2)!} \Delta_{El}$.} and we
used the explicit expression for spin $s$ harmonics with $m=0$
to write them  in terms of associated Legendre polynomials 
(\cite{goldberg}),
\begin{eqnarray}
Y_l^0(\theta,\phi)&=& \sqrt{(2l+1)\over 4\pi} P_l(\cos \theta)
\nonumber \\
\ _{\pm 2}Y_l^0(\theta,\phi)&=& \sqrt{{(2l+1)\over 4\pi}{(l-2)! 
\over (l+2)!}} 
P_l^2(\cos \theta).
\label{eqylm}
\end{eqnarray}
As stated above
for scalar modes in this reference frame $U=0$, so $\Delta_P$
describes both spin $\pm 2$ quantities. 
For $m=0$ one has $\, _2Y_l^0=\,_{-2}Y_l^0$ and so
$ \; _{2}\Delta_{Pl}=\; _{-2}\Delta_{Pl}$.
For density perturbations only $m=0$
harmonics are needed in the expansion because of the azimuthal symmetry,
but the treatment can be 
generalized to vector and tensor modes.
In these  cases modes with $m=\pm 1$
and $m=\pm 2$ are required, respectively. For vector and tensor modes
$U$ is no longer zero, so separate expansions
for both $Q+i U$ and  $Q-i U$ are needed. 

The Boltzmann equation for the CMB photons reads (e.g.
\cite{mabert95}), 
\begin{eqnarray} 
\dot\Delta_T +ik\mu \Delta_T 
&=&-{1\over 6}\dot h-{1\over 6}(\dot h+6\dot\eta)
P_2(\mu) + \dot{\Delta}_{T|Thomson}\nonumber \\   
\dot\Delta_P +ik\mu \Delta_P
&=&\dot{\Delta}_{P|Thomson}.
\label{boltzmann1}
\end{eqnarray}
The first term in the temperature equation represents
the effect of gravitational redshift, while $\dot{\Delta}_{T|Thomson}$
and $\dot{\Delta}_{P|Thomson}$ are the changes in the photon
distribution function produced by Thomson scattering, where the 
derivatives are taken with respect to conformal time.
After inserting equation (\ref{expleg}) into equation (\ref{boltzmann1})
one obtains a system of two coupled hierarchies,
\begin{eqnarray}
\dot{\Delta}_{T0}
 &=& -k\Delta_{T1}-{\dot{h}\over 6}+\dot{\Delta}_{T0|Thomson}
 \, \nonumber \\
\dot{\Delta}_{T1} &=&
{k \over 3}\left[\Delta_{T0}-
2\Delta_{T2}
\right] +\dot{\Delta}_{T1|Thomson}
\,\nonumber\\
\dot{\Delta}_{T2} &=&{k \over 5}\left[2\Delta_{T1}^{(S)}-3
\Delta_{T3}\right] +{2  \over 15} k^2  \alpha+
\dot{\Delta}_{T2|Thomson} \nonumber \\
\dot{\Delta}_{Tl}&=&{k \over
2l+1}\left[l\Delta_{T(l-1)}-(l+1)
\Delta_{T(l+1)}\right]
+\dot{\Delta}_{Tl|Thomson} \,, l>2 \nonumber \\
\; _2\dot\Delta_{Pl}&=&{k\over 2l+1}\left[(l-2)\; _2\Delta_{Pl-1}-
(l+3)\; _2\Delta_{Pl+1}\right]+\; _2\dot{\Delta}_{Pl|Thomson},
\label{eqn1}
\end{eqnarray}
where $\alpha=(\dot{h}+6\dot{\eta})/ 2 k^2$,
and we used the recurrence relations for the Legendre functions,
\begin{eqnarray}
\mu P_l(\mu)={1\over 2l+1}\left[ l\ P_{l-1}+
(l+1) P_{l+1}\right] \nonumber \\
\mu P_l^2(\mu)={1\over 2l+1}\left[(l+2)P_{l-1}^2+
(l-1)P_{l+1}^2\right].
\label{recurr}
\end{eqnarray}

The Thomson scattering cross section is,
\begin{equation}
{ d \sigma \over d \Omega}={3 \sigma_T \over 8 \pi} |\tilde 
\epsilon \cdot \tilde
\epsilon^{\prime}|^2,
\label{thcs}
\end{equation} 
where 
$\tilde \epsilon$ and 
$\tilde{\epsilon^\prime}$
are the unit vectors that describe the polarization of the electric
field
of
the scattered and incoming radiation, respectively. 
The scattering 
terms in equations (\ref{eqn1}) are 
most easily computed in 
the coordinate system where  the incident photons travel
along the $\hat z$ axis and the electrons are at rest.
If  $\hat n^\prime$ is the direction of the incident
photon and $\hat n$ that of the scattered one then
$\hat n^\prime=\hat z=(\theta=0,\phi=0)$ and 
$(\theta,\phi)$ describe $\hat n$. 
For a given scattering event, 
the Thomson scattering matrix is the simplest when
expressed in terms of the
intensities of  radiation parallel ($\tilde T_{\parallel}$) 
and perpendicular ($\tilde T_{\perp}$) to the plane  
containing both $\hat n$ and $\hat n^{\prime}$. 
Equation (\ref{thcs}) leads to
the following relation
between incoming and scattered radiation,
\begin{eqnarray}
\tilde T_{\parallel}
&=&{3\over 8\pi}\sigma_T \cos^2\theta \ \tilde T_{\parallel}^\prime
\nonumber \\  
\tilde T_{\perp}&=&{3\over 8\pi}\sigma_T \ \tilde T_{\perp}^\prime
\nonumber \\  
\tilde U&=&{3\over 8\pi}\sigma_T \cos \theta 
\ \tilde U^\prime,
\label{eqn2}
\end{eqnarray}
where $\sigma_T$ is the Thomson scattering 
cross section.
The total intensity is the sum of the two components, 
$\tilde T=\tilde 
T_{\parallel}+\tilde T_{\perp}$, while the difference gives polarization
$\tilde Q=
\tilde T_{\parallel}-\tilde T_{\perp}$. Because the components are measured
using this coordinate system the Stokes parameters
of the incoming radiation  $\tilde Q^\prime$ 
and 
$\tilde U^\prime$ depend on the angle $\phi$ of the scattered
photon, while $\tilde Q $ and $\tilde U$
are already measured relative to the correct frame.
It is more useful to refer the Stokes parameters of the incoming 
radiation relative to a fixed frame.
To achieve this we  construct the
scattering matrix in terms of  $T^{\prime}$, 
$Q^\prime+iU^\prime
=\exp(2i\phi)(\tilde Q^\prime+i\tilde U^\prime)$
and $Q^\prime-iU^\prime=\exp(-2i\phi)(\tilde Q^\prime-
i\tilde U^\prime)$, where we have used the transformation law 
(equation \ref{QUtrans}) to
relate the two sets of Stokes parameters.

Equation (\ref{eqn2}) implies that 
the scattered radiation in
direction $\hat n$ is
\begin{eqnarray}
\delta T(\hat n^{\prime},\hat n)&=&{\sigma_T \over 4\pi}\left[{3\over 4}(1+\cos^2 \theta) T^\prime +
{3\over 8}(\cos^2 \theta -1)e^{-2i\phi} (Q^\prime+ iU^\prime)+\right.  
\nonumber \\
&& \left.
{3\over 8}(\cos^2 \theta -1)e^{2i\phi} (Q^\prime- iU^\prime)\right] \nonumber \\
\delta (Q\pm iU)(\hat n^{\prime},\hat n)&=&{\sigma_T \over 4\pi}\left[{3\over 4}(\cos^2 \theta-1) T^\prime +
{3\over 8}(\cos \theta \pm 1)^2 e^{-2i\phi}(Q^\prime+ iU^\prime)+ \right. 
\nonumber \\
&& \left.
{3\over 8}(\cos \theta \mp 1)^2e^{2i\phi}(Q^\prime- iU^\prime)
\right]. \nonumber \\
\label{eqna}
\end{eqnarray}
The final expression
for the scattered field is an integral over all directions $\hat
n^\prime$, 
\begin{equation}
\dot X(\hat n)|_{Thomson}
= -a \sigma_T n_e \ x_e  \left[X(\hat n)+
\int d\Omega^{\prime} \delta X(\hat n^{\prime},\hat n)\right],
\label{expfinal}
\end{equation}
where
$X$ stands for $T$ and $(Q\pm iU)$.
The first term 
accounts for the photons that are scattered away from the line of sight and
the expansion factor $a$ is introduced because we are calculating
the derivative with respect to conformal time. 

Equation (\ref{eqna}) for the scattering matrix is written in the
frame where $\hat n^{\prime}=(\theta^{\prime}=0,\phi^{\prime}=0)$.
We can use equation (\ref{eqylm}) to show that 
$_{\pm 2} Y_2^2(\hat n)=\sqrt{1\over
4\pi}(1\mp\cos\theta)^2/2e^{2i\phi}$
and
$_{\pm 2} Y_2^{-2}(\hat n)=\sqrt{1\over
4\pi}(1\pm\cos\theta)^2/2e^{-2i\phi}$. These 
together with 
the explicit expressions  
$_0 Y_0^m(\hat n^{\prime})=\sqrt{1\over 4\pi} \delta_{m0}$,
$_0 Y_2^m(\hat n^{\prime})=\sqrt{5\over 4\pi} \delta_{m0}$,
and $_{\pm 2} Y_2^m(\hat n^{\prime})=\sqrt{5\over 4\pi} \delta_{m\mp
2}$ unable us to rewrite (\ref{eqna}) in a more useful form
($\delta_{ij}$ is the Kronecker delta),
\begin{eqnarray}
\delta T(\hat n^\prime,\hat n)&=&\sigma_T\sum_m\Big[\Big({1\over
10}\ _0Y_2^m(\hat n)\ _0\bar Y_2^{m}(\hat n^\prime)
+\ _0Y_0^m(\hat n) \ _0\bar
Y_0^{m}(\hat n^\prime)  
\Big) T^\prime  
\nonumber \\
& &  -
{3\over 20} \sqrt{2\over 3}\ _0Y_2^m(\hat n)\ _2\bar Y_2^{m}
(\hat n^\prime)\
(Q^\prime+ iU^\prime) -
{3\over 20} \sqrt{2\over 3}\ _0Y_2^m(\hat n)\ _{-2}\bar Y_2^{m}
(\hat n^\prime)
(Q^\prime- iU^\prime)
\Big]  \nonumber \\
\delta (Q\pm i U)(\hat n^\prime,\hat n)&=&\sigma_T\sum_m\Big[-{6\over
20}\ _{\pm 2}Y_2^m(\hat n)\ _0\bar Y_2^{m}(\hat n^\prime)   T^\prime +
{6\over 20} \ _{\pm 2}Y_2^m(\hat n)\ _2\bar Y_2^{m}(\hat n^\prime)
(Q^\prime+ iU^\prime)+ \nonumber \\
&& {6\over 20} 
\ _{\pm 2}Y_2^m(\hat n)\ _{-2} \bar 
Y_2^{m}(\hat n^\prime)(Q^\prime- iU^\prime)
\Big] . 
\label{scm}  
\end{eqnarray}
The advantage of this form for the scattering matrix 
comes from the fact that 
we want the scattering matrix in the frame where $\vec k \parallel
\hat z$ and not $\hat n^{\prime}=\hat z$. 
The sum
$\sum_m\ _sY_l^m(\hat n)\ _{s^\prime}\bar Y_l^{m}(\hat n^\prime)$ 
acquires a phase change under rotation of the coordinate system 
that exactly cancels the phase 
change in the transformation of $(Q\pm i U)$ in equation 
(\ref{scm}), 
we may therefore use this equation 
in the coordinate system where $\vec k \parallel \hat z$ to
compute the Thomson scattering terms in equation (\ref{eqn1}).
Equation (\ref{scm}) is also particularly useful to perform the
integral in equation (\ref{expfinal}).

Substituting the expansion for the Stokes parameters from equation
(\ref{expleg}) into equation (\ref{scm}) and using equation 
(\ref{expfinal})
we find
\begin{eqnarray}
\dot\Delta_{Tl}|_{Thomson}& \equiv & -a \sigma_T n_e x_e \left[\Delta_{Tl}
+ \int d \Omega 
\,_0Y_l^m(\hat n) \delta T(\hat n)\right]
\nonumber \\
&=&\dot \kappa(- \Delta_{Tl}
+\Delta_{T0}\delta_{l0}+{\Pi\over
10}\delta_{l2}) \nonumber \\
\ _{\pm 2}\dot\Delta_{Pl}|_{Thomson}
&\equiv&-a \sigma_T x_en_e \left[\ _{\pm 2}\Delta_{Pl}
+ \int d \Omega \,_2Y_l^m(\hat n)\ \delta(Q\pm iU)(\hat n)\right]
\nonumber \\
&=&\dot \kappa (\ _{\pm 2}\Delta_{Pl}-{\Pi\over
20}\delta_{l2} )
\label{eqnv}
\end{eqnarray}
with $\Pi=\Delta_{T2}-6 (\ _2\Delta_{Pl}+\ _{-2}\Delta_{Pl})$. 
The differential optical depth for Thomson scattering is denoted
$\dot\kappa=an_ex_e\sigma_{T}$. Note that the polarization
has sources  only  at $l=2$. One can apply the same analysis
to vector and
tensor perturbations, the only difference is that in equation (\ref{expleg})
the expansion is in terms of harmonics with 
$m=\pm 1, \pm 2$ for vectors and tensors, respectively. 
In fact, the form of the scattering terms in equation (\ref{eqnv}) is
the same for the three types of perturbations, as also noted by Hu \& White
(1997). 
Equation (\ref{eqnv}) is valid in the rest frame of the electrons,
so in the reference frame where the baryon velocity is $v_b$
the distribution of scattered radiation has an additional 
dipole. 
The final expression for the Boltzmann hierarchy is
\begin{eqnarray}
\dot{\Delta}_{T0}
 &=& -k\Delta_{T1}-{\dot{h}\over 6}
 \, \nonumber \\
\dot{\Delta}_{T1} &=&
{k \over 3}\left[\Delta_{T0}-
2\Delta_{T2}
\right] + \dot{\kappa} \left(
{v_b \over 3}-\Delta_{T1}\right)\,\nonumber\\
\dot{\Delta}_{T2} &=&{k \over 5}\left[2\Delta_{T1}^{(S)}-3
\Delta_{T3}\right] +{2  \over 15} k^2  \alpha+
\dot{\kappa} \left[{\Pi \over 10}-\Delta_{T2}\right]\, \nonumber\\
\dot{\Delta}_{Tl}&=&{k \over
2l+1}\left[l\Delta_{T(l-1)}-(l+1)
\Delta_{T(l+1)}\right]
-\dot{\kappa}\Delta_{Tl} \,, l>2 \nonumber \\
\; _2\dot\Delta_{Pl}&=&{k\over 2l+1}\left[(l-2)\; _2\Delta_{Pl-1}-
(l+3)\; _2\Delta_{Pl+1}\right]-\dot\kappa \; _2\Delta_{Pl}-{1\over 20}
\dot\kappa \Pi \delta_{l2}.
\label{photono}
\end{eqnarray}

\subsection{Non-flat geometry}

We now proceed to generalize the results of the previous section to 
a general Robertson-Walker background.
First we generalize equation (\ref{photono}).
Following Wilson \& Silk (1981) we
expand the photon temperature $\Delta_{T}$ in terms of
Legendre tensors,
\begin{eqnarray}
\Delta_T=\sum_l  (2l+1)
 \; \Delta_{Tl} (-\beta)^{-l} (\prod_l b_l)^{-1} 
G_{|i_1\cdots i_l}\; P_l^{i_1\cdots i_l},  
\label{eqns}
\end{eqnarray}
with $\beta^2=k^2+K$, $b_l^2=1-Kl^2/\beta^2$ and $|i_1\cdots i_l$
covariant derivatives.
These Legendre 
tensors are symmetric combinations constructed out of $\hat n$ 
and the metric $g^{ij}$ using
the recursion
\begin{equation}
(2l+1)\hat n^{(i}\; P_l^{i_1\cdots i_l)}=(l+1) P_{l+1}^{i_1\cdots
i_{l+1}} + l g^{(ii_1}\; P_{l-1}^{i_2\cdots i_l)},
\label{optemp}
\end{equation}
where the parentheses imply symmetrization with respect to the 
indices
(Wilson \& Silk 1981; \cite{white95}). 
The first moment at $l=0$ is given by $P_0= 1$. 
In the flat universe
this recursion reduces to equation (\ref{recurr})
and  equation (\ref{eqns}) is equivalent to equation (\ref{expleg}).

To generalize equation (\ref{expleg}) to polarization 
the Legendre tensors should be constructed in a different way.
Polarization depends both on $\hat n$
and on $(\hat e_1,\hat e_2)$ in the plane 
perpendicular to it.
One may form the linear combination
$\hat{m}=2^{-1/2}(\hat e_1 +i \hat e_2)$ and  
construct the appropriate tensors by combining $\hat n$ and $\hat{m}$.
We expand the polarization
perturbation as
\begin{equation}
\Delta_P=\sum_l  (2l+1)
 \; _2\Delta_{Pl} (-\beta)^{-l} (\prod_l b_l)^{-1} 
G_{|i_1\cdots i_l}\; _2P_l^{i_1\cdots i_l}.  
\end{equation}
These ``spin'' Legendre 
tensors are symmetric combinations constructed out of $\hat n$, 
$\hat{m}$ and the metric $g^{ij}$ using
the recurrence
\begin{equation}
(2l+1)\hat n^{(i}\; _2 P_l^{i_1\cdots i_l)}=(l-1)P_{l+1}^{i_1\cdots
i_{l+1}} +(l+2) g^{(ii_1}\; _2 P_{l-1}^{i_2\cdots i_l)}.
\label{oppol}
\end{equation}
The hierarchy 
begins at $l=2$ with $_2P_2^{ij}=3\ m^i m^j$. 
To derive the Boltzmann hierarchies for polarization 
the following  properties are useful and can be
proven by induction
\begin{eqnarray}
g_{i_1i_2}\; P_l^{i_1\cdots i_l}=0 & \; \;& 
g_{i_1i_2}\; _2P_l^{i_1\cdots i_l}=0 \nonumber \\
\hat n_{i_1}\; P_l^{i_1\cdots i_l}= P_{l-1}^{i_1\cdots i_{l-1}}
&\; \; &
\hat n_{i_1}\; _2P_l^{i_1\cdots i_l}=(l+2)_2P_{l-1}^{i_1\cdots i_{l-1}}/l.
\label{recur2}
\end{eqnarray}
Using equations 
(\ref{optemp}), (\ref{oppol}) and (\ref{recur2}) we  obtain,
\begin{eqnarray}
G_{|i_1\cdots i_li}\hat n^i \; P_l^{i_1\cdots i_l}={1\over
(2l+1)}\left[ -(l+1)G_{|i_1\cdots i_{l+1}}\; P_l^{i_1\cdots
i_{l+1}} + l \beta^2 b_l^2  G_{|i_1\cdots i_{l-1}}\; P_l^{i_1\cdots
i_{l-1}} \right] \nonumber \\
G_{|i_1\cdots i_li}\hat n^i \; _2P_l^{i_1\cdots i_l}={1\over
(2l+1)}\left[ -(l+3)G_{|i_1\cdots i_{l+1}}\; _2P_l^{i_1\cdots
i_{l+1}} + (l-2)\beta^2 b_l^2  G_{|i_1\cdots i_{l-1}}\; _2P_l^{i_1\cdots
i_{l-1}} \right].
\end{eqnarray}
These relations together with
\begin{eqnarray}
\int d\Omega \int dV\ 
(G_{|i_1\cdots i_l}\; P_{l-1}^{i_1\cdots i_l}\
G_{|i_1\cdots i_{l-1}}\; P_{l-1}^{i_1\cdots
i_{l-1}})_{|i}n^i&=&0 \nonumber \\
\int d\Omega \int dV\ 
(G_{|i_1\cdots i_l}\; _{2}P_l^{i_1\cdots i_l}\
G_{|i_1\cdots i_{l-1}}\; _{2}P_l^{i_1\cdots
i_{l-1}})_{|i}n^i&=&0 
\end{eqnarray}
can be used to show that our choice of normalization
in equations (\ref{optemp}) and (\ref{oppol}) coincides with that of 
flat space.

The Boltzmann equation
\begin{eqnarray}
\dot{\Delta}_T+n^i \Delta_{T|i}&=&-{1\over 6}(\dot h + G_{|ij} P_2^{ij})
+ \dot{\Delta}_{T|Thomson} \nonumber \\
\dot{\Delta}_P+n^i \Delta_{P|i}&=&  \dot{\Delta}_{P|Thomson}
\end{eqnarray}
now becomes a hierarchy
\begin{eqnarray}
\dot{\Delta}_{T0}
 &=& -k\Delta_{T1}-{\dot{h}\over 6}
 \, \nonumber\\
\dot{\Delta}_{T1} &=&
{\beta \over 3}\left[ b_1\Delta_{T0}-
2b_2\Delta_{T2}
\right] + \dot{\kappa} \left(
{v_b \over 3}-\Delta_{T1}\right)\,\nonumber\\
\dot{\Delta}_{T2} &=&{\beta \over 5}\left[2b_2\Delta_{T1}-3
b_3\Delta_{T3}\right] +{2  \over 15} k^2 {\bar b} \alpha+
\dot{\kappa} \left[{\Pi \over 10}-\Delta_{T2}\right]\, \nonumber\\
\dot{\Delta}_{Tl}&=&{\beta \over
2l+1}\left[lb_l\Delta_{T(l-1)}-(l+1)b_{l+1}
\Delta_{T(l+1)}\right]
-\dot{\kappa}\Delta_{Tl} \,, l>2 \nonumber \\
\;  _2\dot\Delta_{Pl}&=&
{\beta\over 2l+1}\left[(l-2)b_{l}\; _2\Delta_{Pl-1}-
(l+3)b_{l+1}\; _2\Delta_{Pl+1}\right]-\dot\kappa \; _2\Delta_{Pl}-{1\over 20}
\dot\kappa \Pi \delta_{l2}, 
\label{photono2}
\end{eqnarray}
where
$\Pi=\Delta_{T2}-12 \, _{-2}\Delta_{Pl}$
and ${\bar b}=\sqrt{1-3 K /k^2}=\beta b_2 / k$.  The 
same hierarchy (equation \ref{photono2}) but 
without Thomson scattering and polarization
applies to massless neutrinos, while for massive neutrinos 
the hierarchy depends on the momentum as well (e.g. \cite{mabert95}). 

Finally, the power spectra are given by
\begin{eqnarray}
C_{(T,E)l}&=&(4\pi)^2\int \beta^2d\beta
P(\beta)|\Delta_{(T,E)l}(\beta,\tau=\tau_0)|^2 \nonumber \\
C_{Cl}&=&(4\pi)^2\int \beta^2d\beta
P(\beta)\Delta_{Tl}(\beta,\tau=\tau_0)\Delta_{El}(\beta,\tau=\tau_0)
\nonumber \\
\Delta_{El}&=&-\sqrt{(l+2)! \over (l-2)!} \; _{2}\Delta_{Pl},
\label{cl}
\end{eqnarray}
with $P(\beta)$ denoting the primordial power spectrum. 
Equation (\ref{cl}) only applies to flat and open universes, whereas
for the closed universe
the eigenvalues of the Laplacian are discrete so the 
integral 
is replaced with a sum over $K^{-{1\over 2}} \beta=3,4,5...$  
($K^{-{1\over 2}} \beta=1,2$ are pure gauge modes, Abbott \& Schaefer 1986).
The usual choice for the power spectrum is
\begin{equation}
P(\beta)\propto {(\beta^2-4K)^2 \over \beta (\beta^2-K)}
\end{equation}
which has equal power in the curvature
perturbation per logarithmic interval of $k$ (Lyth \& Stewart 1990, 
White \& Bunn 1995). Note that in equation (\ref{cl}) one integrates
over $\beta$ instead of $k$ as in the flat case.

\section{Integral Solution}

Having derived the Boltzmann equation for temperature and polarization
in the previous section we proceed to derive the integral solution, 
which was the basis of the numerically efficient algorithm in flat 
space (Paper I).
An alternative, more 
geometrical derivation 
of the integral solution is presented in the appendix.  
In the flat case 
the temperature 
and polarization multipoles can be
written as a time integral of the product of a source term  and a 
geometrical term, which is the solution of the source-less
Boltzmann equation. 
The geometrical term (given in terms of
spherical Bessel functions in the flat case) becomes a function 
of two parameters in a non-flat universe, 
because of the additional 
scale in the problem,
the curvature  of the universe. 
The source term
can be expressed as in the flat case 
in terms of the photon, baryon and metric perturbations.
The main property of this solution remains unchanged: 
the source term is a slowly varying function of the 
wavenumber, while the geometrical term, which oscillates much more rapidly,
does not depend on the specific cosmological model except through
its curvature.

To obtain the integral solution it is useful to
work in spherical coordinates. The   
unperturbed metric can be written as 
\begin{equation}
ds^2=a^2 [- d \tau^2+d\chi^2+r^2(\chi)(d\theta^2+\sin^2 \theta d\phi^2)],
\label{metric}
\end{equation}
where the coordinate $\chi$ is related to $r$ by
\begin{eqnarray}
r(\chi)\equiv
\left\{ \begin{array}{ll} K^{-1/2}\sin K^{1/2}\chi,\ K>0\\
\chi, \ K=0\\
(-K)^{-1/2}\sinh (-K)^{1/2}\chi,\ K<0.\\
\end{array}
\right.
\label{rchi}
\end{eqnarray}

In these coordinates the eigenfunctions of the Laplacian 
(equation \ref{Laplace})
are given by
\begin{equation}
G_{sph}(\vec x)=\Phi_\beta^l(\chi)Y_{lm}(\vec n).
\end{equation}
The radial functions $\Phi_\beta^l(\chi)$ are the so called
ultra-spherical Bessel functions. From the expression for the 
Laplacian in spherical coordinates it follows that $\Phi_\beta^l(\chi)$
obey the following differential equation
(Abbott \& Schaefer 1986),
\begin{equation}
{d^2 u_{\beta}^l\over d\chi^2}+\left[\beta^2-{l(l+1)\over r(\chi)^2}\right]
u_{\beta}^l=0,
\label{difeq}
\end{equation}
where $u_{\beta}^l(\chi)=r(\chi)\Phi_{\beta}^l(\chi)$.
In the flat case the solutions for $\Phi_\beta^{l}(\chi)$ reduce to
the familiar spherical Bessel functions $j_l(k\chi)$. The
derivative of an ultra-spherical Bessel function can be written as 
(Abbott \& Schaefer 1986),
\begin{eqnarray}
\dot{\Phi}_\beta^{l}(\chi)={\beta \over 2l+1}\left[l b_l \Phi_\beta^{l}
(\chi)
-(l+1)b_{l+1}\Phi_\beta^{l+1}(\chi)\right],
\label{recrel}
\end{eqnarray}
where the derivative is with respect to $\chi=\tau_0-\tau$, so that 
$d\chi=-d\tau$.
Comparing equations (\ref{photono2}) and (\ref{recrel}) we  see
that the ultra-spherical Bessel
functions  and their derivatives are
solutions of the Boltzmann hierarchy in the absence of scattering and
gravity. 
One can thus make the following ansatz to solve the Boltzmann hierarchy
\begin{equation}
\Delta_{Tl}(\tau)=\int_0^{\tau} d\tau^{\prime} 
e^{-\kappa(\tau,\tau^{\prime})}\left[\Phi_\beta^{l}(\tau-
\tau^{\prime})S_0(\tau^{\prime})
+\dot{\Phi}_\beta^{l}(\tau-\tau^{\prime})S_1(\tau^{\prime})
+\ddot{\Phi}_\beta^{l}(\tau-\tau^{\prime})S_2(\tau^{\prime})\right].
\label{integ}
\end{equation}
Here 
coefficients $S_0,S_1$ and $S_2$ were introduced because there are
sources for $l=0,1,2$ in equation (\ref{photono2}).
By simple substitution of equation (\ref{integ}) in
(\ref{photono2}) we obtain,
\begin{eqnarray} 
S_0&=&\dot{\eta}+\dot{\kappa}\left(\Delta_{T0}+{\Pi \over 4 {\bar b}}
\right)
\nonumber \\
S_1&=&\dot{\kappa}{v_b \over k}\nonumber \\
S_2&=&\alpha+{3 \dot{\kappa}\Pi \over 4 k^2 
{\bar b}}. 
\end{eqnarray}
The values of the ultra-spherical Bessel functions and their
derivatives for $\chi=0$ are also needed in the calculation,
$\Phi_\beta^{l}(0)=\delta_{l0}$, 
$\dot{\Phi}_\beta^{l}(0)={k\over 3} \delta_{l1}$ and 
$\ddot{\Phi}_\beta^{l}(0)={2\over 15} k^2 \bar b \delta_{l2}-
{k^2 \over 3} \delta_{l0}$. These derivatives can be obtains from 
equation $(\ref{recrel})$ using $\Phi_\beta^{l}(0)=\delta_{l0}$.

Finally, integrating equation (\ref{integ}) by parts one can 
eliminate the derivatives of ultra-spherical Bessel functions
and the solution can be written in the following form
\begin{eqnarray}
\Delta_{Tl} &=&\int_0^{\tau_0}d\tau \Phi_\beta^{l}(\tau_0-\tau)
S_{T}(\beta,\tau) 
\nonumber \\
S_T(\beta,\tau)&=&g\left(\Delta_{T0}+2 \dot{\alpha}
+{\dot{v_b} \over k}+{\Pi \over 4 {\bar b}}
+{3\ddot{\Pi}\over 4k^2 {\bar b}}\right)\nonumber \\
&+& e^{-\kappa}(\dot{\eta}+\ddot{\alpha})
+\dot{g}\left({v_b \over k}+\alpha
+{3\dot{\Pi}\over 4k^2 {\bar b}}\right)
+{3 \ddot{g}\Pi \over
4k^2 {\bar b}},
\label{sourceop}
\end{eqnarray}
where we have 
defined the visibility function $g=\dot{\kappa}e^{-\kappa}$.
Comparing this expression with the equivalent one 
for the flat universe (Paper I) one can see that
the only difference in the present case 
is that $\Pi$ is always divided by ${\bar b}$, which comes from the
fact that $\ddot{\Phi}_\beta^{2}(0)={2\over 15} k^2 \bar b$  while
$\ddot{j}_2(0)={2\over 15} k^2$. Thus to transform from the flat to the
general solution one should replace the Bessel functions 
with their non-flat generalization and introduce a factor of 
$\bar b$ to account for
the different normalization of these functions at the origin.

To solve the polarization hierarchy we use the recursion relation
(Abbott \& Schaefer 1986),
\begin{equation}
\sqrt{k^2-K(l+1)^2}\Phi_{\beta}^{l+1}(\chi)=(2l+1)\cot_K(\chi)
\Phi_{\beta}^l(\chi)-\sqrt{k^2-Kl^2}
\Phi_{\beta}^{l-1}(\chi)
\label{recur}
\end{equation}
where 
\begin{eqnarray}
\cot_{K}(\chi)\equiv
\left\{ \begin{array}{ll} K^{1/2}\cot K^{1/2}\chi 
,\ K>0\\
\\
{\chi}^{-1}, \ K=0\\
\\
(-K)^{1/2}\coth(-K)^{1/2}\chi,\ K<0.\\
\end{array}
\right.
\label{cotK}
\end{eqnarray}
Using this and equation (\ref{recrel}) one finds
\begin{eqnarray}
\dot{\left({\Phi_{\beta}^l \over r^2}\right)}=
{\beta\over 2l+1}\left[(l-2)b_{l}{\Phi_{\beta}^{l-1} \over r^2}
-(l+3)b_{l+1}{\Phi_{\beta}^{l+1} \over r^2}\right],
\end{eqnarray}
where the time derivative is taken on $(\Phi_{\beta}^l /r^2)$.
This means that $\Phi_{\beta}^l /r^2$ solves the polarization
free streaming hierarchy equation (\ref{photono2}) and is therefore
the Green's function for polarization. The full solution is 
obtained by a substitution
into equation (\ref{photono2}): 
\begin{equation}
_2\Delta_{Pl}=-\int_0^{\tau_0}d\tau {3g\Pi\over 4} 
{\Phi_\beta^l(\tau_0 -\tau)\over k^2 r^2 \bar b}
\label{integpols}
\end{equation}
An alternative 
derivation of this result and photon transport in general can be found in 
the appendix.

Equations (\ref{sourceop}) and (\ref{integpols}) are  the main results of this
section. 
The numerical implementation of these solutions lead to a 
significant reduction in computational 
time in the flat case. For this to be a successful
strategy we need to show that the ultra-spherical Bessel functions can
be computed efficiently. We turn to this subject next.

\section{The Ultra-Spherical Bessel Functions}

The main difficulty in the numerical implementation of 
equations (\ref{sourceop}) and (\ref{integpols}) 
is the calculation of the 
ultra-spherical Bessel functions $\Phi_{\beta}^l(\chi)$. 
The method used in the flat case was to precompute and tabulate
these functions for all the values of interest. This was feasible
because the functions only depend on the product $k\chi$. In 
the non-flat models another scale is introduced in the problem,
the curvature of the universe. 
The ultra-spherical Bessel functions become functions
of both $\beta$ and $\chi$ separately.
Their tabulation
is not feasible because these functions are rapidly
oscillating in both parameters, and an excessive amount 
of memory would be
needed to store the functions sufficiently densely 
to assure accurate results.
In this section we develop an efficient method 
to compute the functions rapidly and with the necessary accuracy 
for the purpose of CMB calculations.

The simplest approach for calculating the ultra-spherical Bessel
functions would be to use 
the recursion relation in equation (\ref{recur}).
This method is usually recommended as the most efficient if only one
particular 
value of $\Phi_{\beta}^l(\chi)$ is required (e.g. Press et al. 1992). 
Unfortunately, using this recursive method  results in computational time
significantly longer than the time needed to compute the source term, 
so that the total integration time becomes excessive. 
The main drawback of this method
is that for every time step needed to
compute the time integration in equation 
(\ref{sourceop}) all $\Phi_{\beta}^l$ 
are calculated up to the required $l_{\rm max}$. 
However, the smoothness of the $C_l$ spectra 
allows us to calculate them sparsely 
and still obtain very accurate results. 
The structure of the integral solution in equation 
(\ref{sourceop})
does not couple different $l$ modes, and one can use this 
property to significantly reduce computational time (Paper I). 
The second shortcoming of the recursive method is that one does not 
use the value of $\Phi_{\beta}^l$ at the previous time step, although
to guarantee accurate integration of equation
(\ref{sourceop}) one needs to sample $\Phi_{\beta}^l$ sufficiently 
densely that differences between successive values are small.
Both shortcomings suggest that one could efficiently compute 
the ultra-spherical Bessel functions  
directly from the differential equation (\ref{difeq}). Although 
this is not as efficient as the recursive method for any given time
$\tau$, it is much more efficient than the recursive method if one 
requires the values of the ultra-spherical Bessel functions for a number 
of time steps between 0 and $\tau_0$. Moreover, 
one  
only needs to evaluate the functions at the needed values of $l$, which is 
typically every 50th value if 1\% accuracy is desired.

\begin{figure}[t]
\vspace*{9.3 cm}
\caption{ Comparison between ultraspherical Bessel function
$\Phi_\beta^{100}(x)$ and spherical Bessel function $j_{100}(\beta
\sin_K x)$
as a function of $\beta \sin_K x$. The two functions agree qualitatively,
but $\Phi_\beta^{100}(x)$ oscillates much more rapidly and has a higher
amplitude at first oscillation. }
\includegraphics{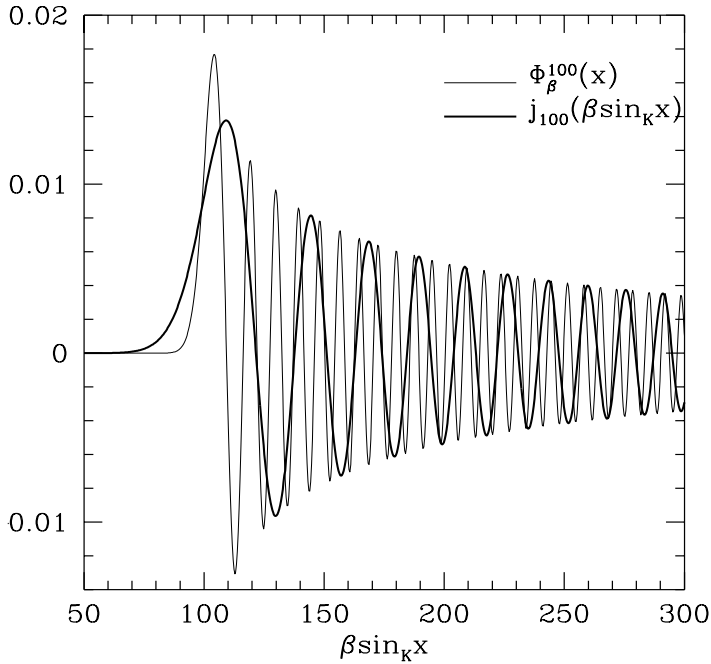}
\label{fig2}
\end{figure}

The structure of the ultra-spherical Bessel functions
is oscillatory and similar to the ordinary Bessel functions (figure 
\ref{fig2}).
The differential equation (\ref{difeq})
can be viewed as the Schr\" odinger equation 
for a particle with
energy $E=\beta^2$ in a potential 
$V=l(l+1) / r(\chi)^2$. Thus for $\chi$ such that
$\beta  r(\chi) > \sqrt{l(l+1)}$, where the ``energy''
is greater than the ``potential'', the solution is oscillatory.
For $\beta r(\chi) < \sqrt{l(l+1)}$ 
there is a growing and a decaying solution, $u_{\beta}^l\propto r^{l+1}$
and $u_{\beta}^l\propto r^{-l}$, respectively.
The growing solution corresponds to $\Phi_{\beta}^l(\chi)$. In order to 
obtain 
an  accurate numerical convergence the integration needs to be started 
in the regime where $\Phi_{\beta}^l(\chi)$ is small, 
which in our case means starting
the integration close to $\beta r(\chi)\approx l$. The equation
then needs to be evolved in the direction of increasing $\chi$  
until recombination, where $\chi \sim \tau_0$. 
The integration of $\Phi_{\beta}^l(\chi)$
therefore proceeds in the opposite time direction than the
evolution of Boltzmann, fluid and Einstein equations (\S 2). 
If one were to start the integration at early times and evolve
the Bessel functions to smaller radial distances $\chi$ (i.e. the  
present time),
the integration would be numerically
unstable, as the decaying $u_{\beta}^l\propto r^{-l}$ mode 
would increasingly contaminate the solution.   

The starting value for the integration of equation (\ref{difeq})
has to be chosen so that $\Phi_{\beta}^l(\chi)$ has some fixed small
value, typically $10^{-6}$. The function is changing rapidly
for $\beta r(\chi) < \sqrt{l(l+1)}$, so one cannot start just anywhere below
this value, because one would soon be in the regime where 
$\Phi_{\beta}^l(\chi)$ is excessively small and numerical round-off
errors become untolerable. One could use 
equation (\ref{recur}) as a downward recursion for a given value of
$\beta$ where $\Phi_{\beta}^l(\chi)$ is of order $10^{-6}$, but this
leads to two difficulties. One is that where $\Phi_{\beta}^l(\chi)=
10^{-6}$ the asymptotic expansion for $\Phi_{\beta}^l(\chi)$
is not necessarily valid as $\chi$ might be too large, 
so starting values of $\chi$ based on the
asymptotic expansion do not necessarily lead to the desired value of
the ultra-spherical Bessel function.
More importantly, using the recursion relation only for one
value of $\chi$ at each $\beta$ still results in an excessive computational
time. The solution is to precompute the starting values for 
integration on a grid of $\beta$ and $l$ and interpolate between them 
for any given value. Because the ultra-spherical Bessel
functions are not oscillatory in this regime accurate interpolation can be  
achieved with only a small number of precomputed values, typically 25 
values of $\beta$  for each $l$. 

One can further reduce the computational 
time by using an asymptotic expansion
for large radial distances $\chi$, which is obtained by  
the WKB approximation applied to equation
(\ref{difeq}), 
\begin{equation}
u_{\beta}^{l}\approx {\sin [\Theta(\chi)] \over 
\beta [\beta^2-l(l+1)/ r(\chi)^2]^{1\over 4}} , 
\label{WKB}
\end{equation}
where 
\begin{equation}
\Theta(\chi)\equiv\int^\chi d\chi \sqrt{\beta^2-l(l+1)/ r(\chi)^2}
\approx \beta \chi+\epsilon .
\end{equation}
Here $\epsilon$ is a 
constant phase that can be obtained analytically,
but we choose it in such a way as
to match the phase at the point where we 
switch from integrating equation (\ref{difeq}) to the WKB approximation
(equation \ref{WKB}). Sufficiently high accuracy is achieved by switching to WKB 
approximation after one hundred  oscillations of
$u_{\beta}^{l}$. 

\section{Approximations}

In this section we compare the results of exact calculations with 
some of the approximations, both those already used in the literature 
and some based on the integral solution. The goal of this section is
to estimate the accuracy of flat space 
approximations, which avoid the need to compute the ultra-spherical 
Bessel functions and were often used
in the context of
reconstruction of cosmological parameters from the CMB. 
To avoid dealing with the complicated structure of ultra-spherical 
Bessel functions one can compute the
$C_l$ spectrum for a flat model and then rescale it in $l$ 
as a result of the change in the angular 
diameter distance to recombination and in the size of
the acoustic horizon
(e.g. \cite{Bond94}, {\cite{Jungman95}, Hu \& White 1996).
The quantities determining the plasma sound horizon and its time 
evolution 
are  $\Omega_b h^2$ and $\Omega_m h^2$, 
with $\Omega_b$ and $\Omega_m$ the density of baryons and matter
in units of critical density.  
If these two parameters are kept fixed the only change in the spectra for
small angular scales should arise from the change in the size of the
acoustic horizon. On large angular scales this approximation fails 
because of additional effects from the decay of the 
gravitational potential
in $\Omega_m \ne 1$ universe
(integrated Sachs-Wolfe contribution), as well as 
curvature effects on the initial 
spectrum and on the radial
eigenmodes.
The simplest and 
most often used approximation is to rescale the spectrum by 
$\Omega_0^{-1/2}$ (e.g. Jungman et al. 1996), while keeping $\Omega_m
h^2$ and $\Omega_b
h^2$ fixed. 
The dependence of the acoustic horizon on $\Omega_0$ is  
only poorly approximated by this scaling and 
leads to significant errors, as shown in figure \ref{fig1} for 
the case $\Omega_0=0.2$. 
Clearly this 
approximation is too inaccurate to be useful for numerical analysis.

Significantly better 
is the approximation where the angular diameter distance to 
the last-scattering surface $r(\chi_r)$ is correctly
calculated and then used to rescale the spectrum with 
(Hu \& White 1996),
\begin{equation}
l=l^{flat}{r(\chi_r) \over \chi_r},
\label{lscale}
\end{equation}
where $\chi_r$ is the radial distance to the recombination epoch.
Recombination occurs roughly around $z \sim 1100$, but 
there is some uncertainty in its exact value and it varies
somewhat with $\Omega_b$. 
The normalization of the spectrum is also a problem, since 
this scaling is only valid for high enough 
values of $l$ making it impossible to use the lower multipoles for
normalization. 
To avoid the two problems 
mentioned above we performed the 
full open calculation and then shifted the flat spectra so that they 
agreed exactly 
at the first peak. This is therefore the most favorable case, 
and the above mentioned difficulties will make the agreement worse.
Figure \ref{fig1} shows this approximation together with the 
exact results. It obviously fails on large scales because of the 
effects mentioned above.
While some of these effects 
can be modeled analytically it is
not obvious how one connects the small scale and large scale 
approximations. 
On small  scales the agreement is significantly better and the 
approximation deviates from the exact calculation by about 3-4\%. 
While this approximation fares 
significantly better than the other approximation above, it is still not
sufficiently accurate for the forthcoming high precision data.

\begin{figure}[t]
\vspace*{9.3 cm}
\caption{ Comparison between the full calculation (solid line),
$\Omega_0^{-1/2}$ rescaling of the flat spectrum (dotted line) and
the angular diameter rescaling of the flat spectrum (dashed line)
for the case with $\Omega_0=0.2$. The latter approximation fares
significantly better than the first one on small scales, while 
on large scales additional effects in open universes make the 
agreement poor for both approximations. }
\includegraphics{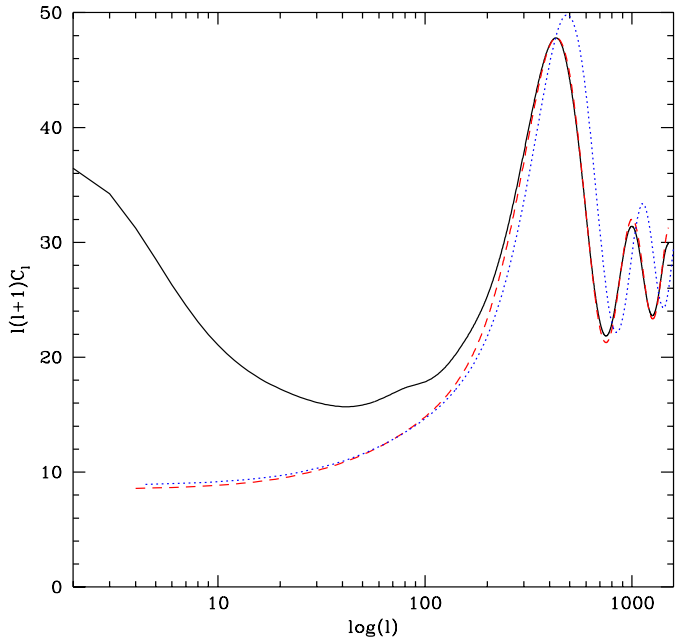}
\label{fig1}
\end{figure}

We also tried 
a more sophisticated version of the above
approximation which solves some of the problems.
In this approximation one uses the integral solution and 
calculates the source term using equations presented in \S2,
but the radial eigenfunctions are replaced with their 
flat space approximations.
This means replacing the ultra-spherical Bessel functions in equation 
(\ref{sourceop})
by the spherical Bessel functions with the argument $kr(\chi)$, so that
the appropriate angular diameter distance relation is used. 
The small scale normalization and epoch of recombination are correctly 
calculated in this case and the integrated Sachs-Wolfe term is included.
If this were a good approximation one could reduce the computational 
time by being able to tabulate the geometrical term as was done
for the flat models (Paper I).
While replacing  the 
ultra-spherical Bessel functions with the flat approximation
gives qualitative agreement with the 
exact results, it {\it does not} converge 
to the exact solution in the limit of large $l$. 
This is shown in figure \ref{fig2}, where 
the two are compared 
for $l=100$. The approximation gives correct qualitative behavior and 
properly describes the position of the
transition from $\Phi_\beta^l(\chi) \approx 0$ to the oscillatory regime, 
as well as the amplitude of oscillations. 
If the last-scattering surface is thin, then in the
time integration of
equation (\ref{sourceop}) 
the spherical Bessel function does not change and it can be taken out of
the integral. Then in equation (\ref{cl}) 
the  square of the Bessel function is multiplied by the square of
the source averaged over time,  this product is 
integrated over the primordial spectrum. Because both 
the primordial spectrum and the
source are slowly changing with $k$, this essentially amounts to an 
averaging of $j_l^2$ over its oscillation period. The result
only depends on the amplitude and not 
on the oscillation frequency of the spherical Bessel function. 
This is the underlying reason why the approximation of rescaling 
the distance to the last-scattering surface
gives qualitative agreement in the final spectrum and
in particular, it correctly predicts
the positions of the acoustic oscillation peaks.

However, as shown in figure \ref{fig2} 
the frequency of oscillation is different for the two
functions, and this has important consequences when the finite width
of the last scattering surface is taken into account. In this case
one has to explicitly
integrate the sources over the recombination epoch as in 
equation (\ref{sourceop}).
The integrated Sachs-Wolfe
term cannot be calculated correctly using this approximation because the
time dependence of the Bessel functions is important. 
This disagreement is most severe for the lower multipoles.
The agreement for the
part proportional to the visibility function is not good either, because
$j_l[\beta\sinh(\chi)]$ oscillates faster in $\chi$
than $\Phi_{\beta}^l$
(the phase is proportional to 
$\beta \sinh(\chi)$ rather than to $\beta \chi$), so  the 
damping is more severe than in its ultra-spherical counterpart. 
The situation is 
even worse for the terms proportional to the first and second derivatives 
of the visibility function (equation \ref{sourceop}). 
In the limit where the visibility function can
be approximated by a $\delta$-function this is easy to understand. The 
derivatives of the visibility function act as derivatives of
the $\delta$-function so these terms 
involve derivatives of the ultra-spherical Bessel functions, 
which are not well approximated by the derivatives of 
$j_l(\beta\sinh \chi)$. This can be seen from the fact that the latter,
although having the same amplitude as $\Phi_{\beta}^l$,
oscillate more rapidly. This problem 
could be solved by using equation (\ref{recrel}) to express 
$\dot\Phi_{\beta}^l$
in terms of the $\Phi_{\beta}^{l-1}$ and $\Phi_{\beta}^{l+1}$
and approximating these by their flat counterparts. But the rest of
the problems mentioned above will remain.
This flat space approximation is therefore 
not suitable for the integral approach and its results are even worse
than the simple rescaling of the spectrum with the angular diameter 
distance. To summarize the discussion, 
none of the flat space approximations can match the exact 
results at better than 5\% accuracy, which is not sufficient for the
expected future observational precision.

\section{Implementation} 

We are now in a position to present an efficient implementation of the
method developed in the previous sections. Some of the technical issues
of this implementation are 
identical to those in the flat case (Paper I) and will not be repeated
here, but non-flat 
geometry also introduces some additional complications.
The temperature spectra are calculated by integrating
equation
(\ref{sourceop}), while the source in this equation 
is obtained 
by integrating the equations presented in \S2. One of the advantages of
calculating the $C_l$ spectra this way is that the source in the integral
only depends on the first few photon multipoles, so accurate results
can be obtained truncating the Boltzmann
hierarchy at a very low value of $l$. As discussed in Paper I this is
only possible if one uses a non-reflective boundary conditions to 
close the hierarchy in equation (\ref{photono2}), so that the propagation of 
power from large scales to small scales is not reflected back onto the
lowest moments of photon distribution. In the absence of scattering and
integrated Sachs-Wolfe contribution equation (\ref{photono2}) 
becomes identical to equation 
(\ref{recrel}), hence one can use the recursion relation in equation 
(\ref{recur})
to relate $\Delta_{l+1}$ to $\Delta_{l}$ and thus close the 
hierarchy. In reality this relation is not exact, but 
since only the lowest moments need to be accurately calculated,
one can use this closure scheme at some moment $l_0$, which is 
sufficiently high so that 
any reflection of power from $l_0$ down to the lowest moments will
be negligible. In practice this is achieved to a high accuracy
for $l_0=7$, 
similar to the flat case.
This therefore 
reduces the number of equations needed to be solved from a few thousand
in the standard Boltzmann code to a few dozen in the integral 
approach, making the calculations significantly faster.

It was mentioned in the previous section that the sources and the 
ultra-spherical Bessel functions are calculated in the opposite directions
of time, hence the integral in equation (\ref{sourceop})
cannot be computed at the same time as the sources are computed. 
The radial eigenfunctions in equation (\ref{sourceop})
are calculated from equation (\ref{difeq}) using a finite differencing
scheme such as Runge-Kutta 4th order integration, a procedure that
automatically produces sufficient
sampling points in time for an accurate integration. Because the sources
do not depend on $l$ and as discussed below 
need to be computed at far fewer 
points than the radial eigenfunctions, we compute them first at 
selected values of wavevector $k$ by solving the system of equations
presented in \S2.
Their values are stored at selected time intervals. During the
integration of equation (\ref{difeq}) the sources are 
interpolated using cubic splines from their tabulated values and
equation (\ref{sourceop}) is solved simultaneously. 
To obtain an accurate interpolation of
the sources, at most a few hundred values in time are needed, 
covering the time interval from recombination until today.

The main advantage of the integral method is in the sampling 
of the wavenumber $k$ for the source term. 
The sources vary typically on a scale
$k\sim 1/ \tau_r$ where $\tau_r$ is the conformal time of recombination.
The oscillations in $\Phi_{\beta}^l$ are approximately $50$
times  faster, as they occur on a scale of the order of the inverse
of the present conformal time, $\tau_0$. We take advantage of this
by calculating the sources at fewer values of $k$ than what is
needed to properly sample the variations of the radial eigenfunctions,
and then interpolate them to obtain their values for
the rest of the values of $k$. This reduces 
the number of $k$ modes where the system of equations in \S2 is
solved, and is an important factor that enhances the 
performance of the method. 

Because 
the $C_l$ spectra are so smooth, they can be sampled in every 50th $l$,
except for
low values of $l$ where denser sampling is needed. The rest of the
$C_l$'s are obtained by interpolation. This reduces the number of 
radial eigenfunctions that need to be computed, so that approximately
45 values of $l$ are needed up to $l_{\rm max}=1500$. They need
to be sampled at least at a few
points per oscillation, resulting in approximately 2000 points
for the above $l_{\rm max}$. One therefore needs to solve $10^5$
differential equations for the radial eigenmodes and only about 3000 
differential equations for the source term. However, the
differential 
equations for the sources are more complicated, so the actual 
computational time
for the two terms is comparable. Typical 
CPU time for an open model with 1\% accuracy up to $l_{\rm max}=1500$
is about 60 seconds on an ALPHA-500 workstation, which 
is about two times slower than for a comparable model in a 
flat universe and orders of magnitude faster than any other 
method.

\section{Discussion}

We derived the photon Boltzmann hierarchy
for scalar temperature and polarization perturbations 
in a general Robertson-Walker 
universe using a new expansion method in spin-$s$ harmonics. Recently,
a similar expansion has been presented by Hu \& White (1997) and applied
to scalar, vector and tensor harmonics, but only 
in a flat universe. Combining 
these techniques will provide 
a complete treatment of cosmological perturbation theory in a 
perturbed Robertson-Walker universe and will be presented elsewhere
(Hu et al. 1997). In this paper we concentrate on providing an efficient
method to compute scalar CMB temperature and polarization anisotropies in 
a non-flat universe, generalizing the methods presented in Paper I.
For this purpose we present
an integral solution to the Boltzmann hierarchy
equations. 
Temperature and polarization perturbations are written
using Green's method
as a time integral over the 
source term multiplied by a radial eigenfunction. 
The first term depends
on the cosmological model, but not on the multipole moment,
while the latter depends only on the curvature of space, but not 
on the rest of the 
cosmological parameters. This split clearly separates
between effects that depend on the source and those that depend on geometry
and is more physically intuitive than the differential formulation. 
The numerical implementation of the integral
method requires an efficient and
accurate way of calculating the ultra spherical Bessel functions, 
because these cannot be stored in advance as in the flat case. 
We achieve this by solving their differential equation. 
The implementation of the integral 
solution leads to a fast and accurate method for 
computing CMB anisotropy and polarization spectra with computational
times of the same order as in the flat case. It should be noted that
once the ultra spherical Bessel functions are calculated computing
vector and tensor spectra poses no further calculational difficulties.

The integral method
presented in this paper can be
used for studying theoretical predictions from various cosmological 
models with the purpose of comparing them to 
the real data. 
The implementation of the method is publicly available. 
The output of the calculation consists of both the
temperature and polarization spectra as well as
the matter density power spectra, so
one can use
it to analyze both clustering and CMB data. In addition, 
an accurate small scale normalization ($\sigma_8$) to the COBE data
is provided by using the 4 year analysis of the 
power spectrum (Bunn \& White 1997).
As such it should replace various 
approximative methods used in the literature if high precision 
is required.
A new generation of experiments
will be able to determine both the density power spectrum (e.g. 
Sloan Digital Sky Survey and the 2dF survey)
and the CMB spectrum (MAP, Planck) to an exquisite accuracy. 
The integral 
method presented here provides a tool for rapid and highly accurate 
analysis of theoretical models over a large range of parameter space.
This should be useful in 
achieving a higher level of precision in the determination of 
the true cosmological model.

\acknowledgements
We acknowledge useful discussions with W. Hu, M. Mahachek and
M. White. We also want to thank the referee, Douglas Scott for many
useful comments.
This work was partially supported by grant NASA  NHG5-2816.

\appendix
\section{Appendix}

In this appendix we present an alternative and more intuitive derivation
of the integral solution. We first present a 
solution of the photon free
streaming equation and then construct the general solution from it 
using Green's method. This method clarifies the structure
of the radial part of the integral solution and the connection between 
spatially varying perturbations and the resulting
angular variation in the sky.

Polarization is described by $Q\pm iU$ which have spin
$s=\pm 2$. The fields $T$, $Q\pm i U$ are function of position in
space specified in spherical coordinates by $(\chi,\theta,\phi)$ as
well as direction in the sky.
We take the observer to be located at the origin, in 
spherical coordinates at $\chi=0$ so we are interested in 
$T$ and $Q \pm iU$ as a function of the direction of observation
$\hat n$ at the origin today. As discussed in the main text
polarization is not only a function of $\hat n$ but also of the
two directions perpendicular to $\hat n$, $(\hat e_1, \hat e_2)$, 
which are used to define the Stokes parameters. Thus,
\begin{eqnarray}
T&=&T(\chi,\theta,\phi,\hat n,\tau)\nonumber \\   
Q \pm iU&=&(Q\pm iU)(\chi,\theta,\phi,\hat n, \hat e_1,\hat
e_2,\tau) 
\end{eqnarray}
We can make a single vector out of these quantities 
${\cal X}=(T,Q+iU,Q-iU)$.

We begin by considering the solution to the free streaming equation,
\begin{equation}
{d {\cal X} \over d\tau}=0
\end{equation}
and then construct the complete solution from that.
Notice that this is only true if the polarization vector
does not rotate relative to  $(\hat e_\theta,\hat e_\phi)$ during
propagation, which can
happen in some Bianchi models (\cite{tolman}). 
Given the field ${\cal X}^*(\chi,\theta,\phi,\hat n, \hat e_1,\hat
e_2,\tau^*)\equiv {\cal X}^*$ at some initial time $\tau^*$ the
free streaming solution for the field at the origin today is simply
\begin{equation} 
{\cal X}(0,\hat n, \hat e_\theta,\hat
e_\phi,\tau_0)={\cal X}^*(\chi^*,\theta,\phi,\hat n, \hat e_\theta,\hat
e_\phi,\tau^*)
\label{freestr}
\end{equation}
where $(\theta,\phi)$ specify the direction of $\hat n$ in spherical 
coordinates and we use $(\hat e_\theta,\hat e_\phi)=(\hat e_1,\hat e_2)$.
Because photon position and time are related via $\chi^*=\tau_0-\tau^*$ 
we may drop the explicit dependence on $\tau^*$ in the following.
When considering the perturbations produced by one single scalar mode,
$G(\vec x)$, the
complete initial field ${\cal X}^*=(T^*,Q^*+iU^*,Q^*-iU^*)$ must be
constructed from that scalar mode. The temperature is a spin zero
function of $\hat n$ and so $G$  and its radial derivatives
can be used to construct the most general form of $\cal X^*$ for 
this mode,
\begin{equation}
\Delta T^*=\ _0\epsilon_0 G+\ _0\epsilon_1 G_{|i}\hat n^{i} + \ldots
\label{texp}
\end{equation}
where we introduced the coefficients $_0 \epsilon_i$ to expand the 
temperature perturbation in terms of the mode functions and their
derivatives.  
We concentrate on the first term, which corresponds to an initial 
distribution that is spatially varying as $G$ and
locally isotropic. It is through the free streaming of the 
photons that the spatial variations of the source create an angular
dependent distribution function despite being isotropic initially.
Taking $G$ as the eigenfunction of
the Laplacian in spherical coordinates
$G_{spher}=\Phi_\beta^l(\chi)Y_{lm}(\theta,\phi)$ (\cite{abbott86})
we find using equation (\ref{freestr}),
\begin{equation}
\Delta T(0,\hat n,\tau_0)=\ _0\epsilon_0 \Phi_\beta^l(\tau_0-\tau^*)
Y_{lm}(\theta,\phi)
\end{equation}
This means that $\Phi_\beta^l(\tau_0-\tau^*)$ is a solution of the
free streaming equation, which we used in the main text
to construct the full solution of the Boltzmann equation for 
temperature. The source terms for temperature consist
not only of isotropic term, but also of higher multipoles. 
Therefore, for the complete solution we have to include also the terms
proportional to the derivatives of the ultra-spherical Bessel
functions, corresponding
to  
higher order terms in the expansion of equation (\ref{texp}).

For polarization we need an initial field
of spin 2, which is a function of
$\hat n$ and $(\hat e_\theta,\hat e_\phi)$. 
We can construct a function
like this by contracting the 
derivatives of $G$ with $\hat m=2^{-1/2}(\hat e_\theta + i\hat
e_\phi)$, $\hat{\bar m}=2^{-1/2}(\hat e_\theta - i\hat e_\phi)$ 
and $\hat n$ in the following way,
\begin{eqnarray}
\Delta (Q+iU)^*&=&_2\epsilon_0\ G_{|ij}\hat m^i \hat m^j+\ _2\epsilon_1
G_{|ijk}\hat m^i \hat m^j\hat n^{k} + 
\ldots \nonumber \\
\Delta(Q-i U)^*&=&_{-2}\epsilon_0\ G_{|ij}\hat{\bar m}^i\hat{\bar m}^j+\
_{-2}\epsilon_1 G_{|ijk} \hat{\bar 
m}^i \hat{\bar mi}^j \hat n^{k} +
\ldots 
\end{eqnarray}
where again we have introduced expansion coefficients $_{\pm 2}
\epsilon_i$.
For polarization we only need to consider the first term in each
expansion because this
suffices for the angular structure of the source.
The free streaming solution is 
\begin{eqnarray}
\Delta (Q+i U)(0,\hat n, \hat e_\theta,\hat
e_\phi)&=&\ _2\epsilon_0\ G_{|ij}\hat m^i \hat m^j
\nonumber \\
\Delta (Q-i  U)(0,\hat n, \hat e_\theta,\hat
e_\phi)&=& \ _{-2}\epsilon_0\ G_{|ij}
\hat{\bar m}^i\hat{\bar m}^j. 
\label{fs}
\end{eqnarray}

By taking derivatives of $G_{spher}(\vec x)$ we can show that
\begin{eqnarray}
G_{spher|ij}\hat m^i \hat m^j
&=&\sqrt{(l+2)!/(l-2)!}{\Phi_\beta^l(\chi) \over
r(\chi)^2}\; _2Y_{lm}(\theta,\phi) \nonumber \\
G_{spher|ij}\hat{\bar m}^i\hat{\bar m}^j
&=&\sqrt{(l+2)!/(l-2)!}{\Phi_\beta^l(\chi) \over
r(\chi)^2}\; _{-2}Y_{lm}(\theta,\phi) 
\end{eqnarray}
so that equation (\ref{fs}) becomes
\begin{eqnarray}
\Delta (Q+i U)(0,\hat n, \hat e_\theta,\hat
e_\phi)&=&\ _2\epsilon_0
\sqrt{(l+2)!/(l-2)!}{\Phi_\beta^l(\chi) 
\over
r(\chi)^2}\; _2Y_{lm}(\theta,\phi) \nonumber \\ 
\Delta (Q-i U)(0,\hat n, \hat e_\theta,\hat
e_\phi)&=&\ _{-2}\epsilon_0 \sqrt{(l+2)!/(l-2)!}{\Phi_\beta^l(\chi) \over
r(\chi)^2}\; _{-2}Y_{lm}(\theta,\phi),
\label{freestsol}
\end{eqnarray}
with $\chi=\tau_0-\tau$.
This shows that  ${\Phi_\beta^l /
r(\chi)^2}$ is the free streaming solution for polarization as 
was obtained using a different method in the main text. 

The complete Boltzmann equation for polarization is of the form
\begin{eqnarray}
{d\Delta (Q+iU)\over d\tau}&=&-\dot\kappa \Delta (Q+iU)
-{3 \dot\kappa \Pi \over 2 k^2 \bar b} 
G_{|ij}\hat m^i \hat m^j \nonumber \\ 
{d\Delta (Q-iU)\over d\tau}&=&-\dot\kappa \Delta (Q-i U)
-{3 \dot\kappa \Pi \over 2 k^2 \bar b} 
G_{|ij}\hat{\bar m}^i \hat{\bar m}^j 
\label{fullbol}
\end{eqnarray}
The
$G_{|ij} m^i m^j$ leads to the usual
$(1-\mu^2)\propto(1-P_2(\mu))$ dependence of Thomson scattering.
The angular dependence of the source is the simplest spin two function
that can be constructed from a scalar. One can interpret the above
equation as stating that at each time $\tau^*$ a contribution
$d(\Delta Q+ i \Delta U) \propto
\dot\kappa \Pi G_{|ij}m^i m^j d\tau$ is generated and then 
free streamed until today. If there is scattering along the 
way only a fraction $\exp(-\kappa)$ of the photons reach the
observer. Given the free streaming solution (equation 
\ref{freestsol}) the complete solution can be written as 
\begin{eqnarray}
\Delta (Q+i U)(0,\hat n, \hat e_\theta,\hat
e_\phi,\tau_0)&=& - \sqrt{(l+2)!/(l-2)!}\left[\int_0^{\tau_0}d\tau g(\tau)
{3 \Pi \Phi_\beta^l(\tau_0-\tau) \over 4
k^2 \bar b r(\chi)^2}\right] \; _2Y_{lm}(\theta,\phi) \nonumber \\ 
\Delta (Q-i U)(0,\hat n, \hat e_\theta,\hat
e_\phi,\tau_0)&=&-
\sqrt{(l+2)!/(l-2)!}\left[\int_0^{\tau_0}g(\tau) \epsilon
{3 \Pi \Phi_\beta^l(\tau_0-\tau) \over 4 k^2 \bar b
r(\chi)^2}\right] \; _{-2}Y_{lm}(\theta,\phi) \nonumber \\
\label{fullbolsol}
\end{eqnarray}
where $g(\tau)$ is the visibility function. When considering 
solutions of equation (\ref{photono2}) 
corresponding to one mode that is the generalization of a single plane
wave a superposition of solutions (\ref{fullbolsol}) 
for all $l$ with $m=0$ must be used. 

The above solutions for temperature and polarization
help to illustrate the
relation between the radial part of the mode function $G$ and
the free streaming solution for $\cal X$ as a function of angle in the
sky. When we look in a given direction on the sky we observe
what was happening away from us at a distance $\chi=\tau_0-\tau^*$. This
couples the spatial variation of the mode functions with the angular
variations of the observed distribution. The angular dependence of the
sources, which is constrained by the spin nature of the 
variables, is also responsible for the specific form of the
integral solution. The source for scalar polarization has to
be constructed out of derivatives of the mode functions which change the
radial dependence of the solution from $\Phi_\beta^l$ to $\Phi_\beta^l/r^2$.  

These arguments can be extended
to vector and tensor modes.
The radial and angular dependence of the mode functions
differs from the scalar case and can be found in Tomita (1982).
In the case of temperature the source is proportional to 
$h_{rr}=h_{ij}\hat n^i \hat n^j$,
while for polarization it is proportional to 
$h_{ij} \hat m^i \hat m^j$ and $h_{ij} \hat{\bar m}^i  \hat{\bar m}^j$ for
$Q+iU$ and $Q-iU$ for any type of perturbations. 
This together with the explicit mode functions
(Tomita 1982) can be used to verify the flat 
space solution obtained previously (Zaldarriaga \& Seljak 1997).
Detailed analysis of vector and tensor non-flat Boltzmann equation 
will be presented elsewhere (Hu et al. 1997).

\end{document}